\documentstyle[12pt]{article}

\newcommand{\be}{\begin{equation}}
\newcommand{\ee}{\end{equation}}
\newcommand{\bn}{\begin{eqnarray}}
\newcommand{\en}{\end{eqnarray}}

\begin{document}

\begin{center}

\noindent{\large\bf
{\LARGE On the dual equivalence of the self-dual and topologically massive p-form models}
}
\vspace{3mm}

\noindent{
{\Large R. Menezes$^1$, J. R. S. Nascimento$^1$, R. F. Ribeiro$^1$, and C. Wotzasek$^{2}$}
}\vspace{3mm}

\noindent

\begin{large}
{\it $^1$Departamento de F\' \i sica, Universidade Federal
da Para\'\i ba\\ 58051-970 Jo\~ao Pessoa, Para\'\i ba, Brasil. \\
$^2$Instituto de F\'\i sica, Universidade Federal do Rio de Janeiro\\
21941-972, Rio de Janeiro, Brazil. } \\
\end{large}
\end{center}

\begin{abstract}
\noindent We study the duality symmetry in p-form models containing a generalized $B_q\wedge F_{p+1}$ term in spacetime manifolds of arbitrary dimensions. The equivalence between the $B_q\wedge F_{p+1}$ self-dual ($SD_{B\wedge F}$) and the
$B_q\wedge F_{p+1}$ topologically massive ($TM_{B\wedge F}$) models is established using a gauge embedding procedure, including the minimal coupling to conserved charged matter current. The minimal coupling adopted for both tensor fields
in the self-dual representation is transformed into a non minimal magnetic like coupling in the
topologically massive representation but with the currents swapped. It is known that to establish
this equivalence a current-current interaction term is needed to render the matter sector unchanged.  We show that both terms arise naturally from the embedding adopted. Comparison with Higgs/Julia-Toulouse duality is established.\\
\end{abstract}

\section{Introduction}

Antisymmetric tensors (p-form fields) are largely used in Physics. They are naturally used to extend the usual four dimensional phenomena to other dimensions. An abelian antisymmetric tensor potential was probably first used in the context of the
particle theory to describe a massless particle of zero-helicity \cite{OP,SD}.  It reappeared later on in the context
of fundamental strings \cite{KR,RW}, has been used to study cosmic strings \cite{VV,DS,RLD} and to put topological charge (hair) on black holes\cite{MJB,ABL,BAC}. The free theory of a rank-2 antisymmetric tensor has also been intensively studied both classically \cite{RKK} and quantically \cite{BK,MO} and has been shown to be dynamically dual (under the Hodge mapping) to a massless scalar field (zero-form). 
In condensed matter physics tensor order parameters appear in the context of dual formulation of the London limit of the Ginzburg-Landau action \cite{GL1}  and in $\mbox{}^3$He-A systems \cite{HEA}. Nowadays p-form fields are largely used in cosmological models in the context of string/brane theories.

The main goal of this work is to study duality symmetry in the context of the p-form models presenting a $B \wedge F$-like term, viz., in models presenting a topological, first-order derivative coupling between forms of different ranks.
These models naturally display the dimensional extension of the duality between the
self-dual (SD) \cite{TPvN} and Maxwell-Chern-Simons models (MCS) \cite{DJT}, shown by Deser and
Jackiw \cite{DJ} or the four dimensional $B_2 \wedge F_2$ model \cite{BF}. The existence of a constraint of self duality in the massive, non invariant model ($SD_{B\wedge F}$) will be established. To this end a new dynamical embedding formalism \cite{IW,AINRW}, that is alternative to
the master Lagrangian approach, will be adopted to obtain the gauge invariant $B_q\wedge F_{p+1}$ model.

Duality is a ubiquitous symmetry concept. It displays the connection of two opposite regimes for the same dynamics playing an important role in nowadays physics,
both in the original contexts of condensed matter and Maxwell electromagnetism, as well as
in the recent research of extended objects. The existence of such a
symmetry within a model has important consequences - it can
be used to derive (exact) non perturbative results since swapping opposite
regimes allows a perturbative investigation of theories
with large coupling constants.

The study of this symmetry has received renewed interest in recent research in diverse areas in
field theory such as, supersymmetric gauge theories \cite{SW},
sine-Gordon model \cite{DC}, statistical systems \cite{JLC} and, in the
context of condensed matter models, applied for instance to planar high-T$_C$
materials, Josephson junction arrays \cite{DST} and Quantum Hall
Effect \cite{MS}. In particular the duality mapping has been of great significance in order to extend the bosonization
program from two to three dimensions with important phenomenological
consequences \cite{boson}. It also plays preponderant role in the ADS/CFT correspondence \cite{maldacena}
that illustrates the holographic principle \cite{t'hooft}.

The idea of duality has also been used in recent developments of  string
theory \cite{EK}, where different vacua are shown to be related by duality\cite{SZ}.  In this context a general procedure for constructing dual models was proposed by Busher \cite{THB} and generalized by Rocek and Verlind \cite{RV} that consists in lifting the global symmetry of the tensor fields with a new gauge field, whose field strength is then constrained to zero by the use of a Lagrange multiplier.  Integrating, sequentially, the multiplier and the gauge field yields the original action while the dual action is obtained if one integrates the gauge field together with the original tensor field, keeping the Lagrange multiplier that then plays the role of dual field to the original tensor field. This line of research was used in the investigation of bosonization as duality by Burgess and Quevedo \cite{BQ} and to discuss S-duality, the relation between strong and weak couplings in gauge theories \cite{EW}.  This procedure has also been shown to be related to canonical transformations \cite{AAGL}. Recently, this line of research has been applied in the context of the topologically massive $B_2\wedge F_{2}$ theory, which is related to our interest here, to study its equivalence with the Stuckelberg construction of gauge invariant massive excitations \cite{SH}.

In this work we deal with the dual equivalence between models describing the same physical phenomenon involving the presence of a topological term in a spacetime of arbitrary dimension.
It is closely related to, (i) the odd-dimensional duality involving the Chern-Simons term (CST) \cite{CST} whose paradigm is the equivalence between SD \cite{TPvN} and MCS \cite{DJT,DJ} theories in (2+1) dimensions and, (ii) to the even dimensional $B \wedge F$ model widely used in 
four-dimensional models, as an alternative mass generation mechanism to the Higgs phenomenom.  
As shown in \cite{DJ}, in three dimensions there are two different ways to describe the dynamics of a single, freely propagating spin one massive mode, using either the SD theory \cite{TPvN} or the MCS theory.
The identification that relates the basic field of the SD model with the dual of the MCS field has been established \cite{DJ}. This correspondence displays the way the gauge symmetry of the MCS representation, gets hidden in the SD representation \cite{DJ}.
It is the presence of the topological and gauge invariant Chern-Simons term the responsible for the essential features manifested by the three dimensional field theories, while in the four dimensional context this role is played by the $B_2\wedge F_{2}$ term.
To extend this duality symmetry relation and study its consequences in the context of field theories in Minkowski manifolds of arbitrary dimensions, including the presence of a $B_q\wedge F_{p+1}$ term, is the main purpose of this work.

This paper is organized as follows. In the next section, we investigate the gauge non invariant $SD_{B\wedge F}$ model, define a new, non-Hodge, (derivative) duality operation  and show the existence of self-duality.  Next, in Section III, we apply an iterative dynamical embedding procedure to construct an invariant theory out of the self-dual $B_q\wedge F_{p+1}$ model - the topologically massive $B_q\wedge F_{p+1}$ model ($TM_{B\wedge F}$).  This is a gauge embedding procedure that is done with the inclusion of counter
terms in the non invariant action, built with powers of the Euler tensors (whose kernels give the field equations for the potentials $A_p$ and $B_q$) to warrant the dynamical equivalence.
Such construction discloses hidden
gauge symmetries in such systems.
The minimal coupling with external matter current and its consequences are studied. We also consider, at the end of the section, the comparison of this duality with the results of \cite{QT,QT2}. Our results are discussed in the final section of the paper.

\section{Self Dual $B_q\wedge F_{p+1}$ Theory}

The study of gauge theories with a topological term, in (3+1) dimensions, has received considerable attention recently.  Among other possibilities the $B_2\wedge F_{2}$ term is interesting for providing a gauge invariant mechanism to give mass to the gauge field and to produce statistical transmutation in (3+1) dimensions. This mechanism may be dimensionally extended straightforwardly.
In $D=4$, it displays a Kalb-Ramond field $B_2$, i.e., a totally antisymmetric tensor potential (a potential 2-form) while $F_2 = dA_1$ is the field strength of
the one-form potential $A_1$. In arbitrary dimensions $B_q$ is a q-form field while $A_p$ is a p-form such that $p+q = D-1$. In this context $F_{p+1} = d A_p $ is the field strength of the p-form potential.

The model with a built-in SD constraint in (2+1) dimensions was proposed in \cite{TPvN} as an alternative to
the concept of topologically massive modes proposed in \cite{DJT}. The former is a non gauge
invariant, first order model, while the later is a second order gauge invariant
formulation, both making use of the topological Chern-Simons term.  In this section
we want to formulate and study a D-dimensional first order, non gauge invariant model, making use of the
topological $B_q\wedge F_{p+1}$ term and prove the existence of self duality property as a consequence of a built in
SD constraint.

The model in question shows the coupling of a p-form field potential $A_p$
with a rank-q tensor field potential $B_q$ \cite{ABL} as,
\begin{equation}
\label{PB05}
{\cal L}_{SD}^{(0)} = \frac 12 (-1)^{p+1}m^2 A_{p}^2 + \frac 12 (-1)^{q+1}m^2 B_{q}^2
+\frac {\chi\,\theta}{2} m  B_{q} \epsilon \partial A_{p} \; .
\end{equation}
The appearance of two parameters, $m$ and $\theta$ in the theory is however illusory.  After a scaling  redefinition as 
\begin{eqnarray}
x &\to & m^{-1} x\nonumber\\
A_p &\to & m^{\frac {D-2}2} A_p\nonumber\\
B_p &\to & m^{\frac {D-2}2} B_p
\end{eqnarray}
to work with dimensionless variables we obtain
\begin{equation}
\label{PB10}
{\cal L}_{SD}^{(0)} = \frac 12 (-1)^{p+1} A_{p}^2 + \frac 12 (-1)^{q+1} B_{q}^2
+\frac {\chi\,\theta}{2}   B_{q} \epsilon \partial A_{p} \; ,
\end{equation}
keeping only the dimensionless coupling constant $\theta$ in the $B_q\wedge F_{p+1}$ term.
Here $A_p \equiv A_{\mu_1 \ldots \mu_p}$, $A_p^2 \equiv A_{\mu_1 \ldots \mu_p}A^{\mu_1 \ldots \mu_p}$ and $B_{q} \epsilon \partial A_{p} \equiv B_{\mu_1\ldots\mu_q}\epsilon^{\mu_1\ldots\mu_q \alpha\nu_1\ldots\nu_p}\partial_\alpha A_{\mu_1 \ldots \mu_p}$. The superscript index in the Lagrangian is the counter of the iterative algorithm
to be implemented in the sequel, $\chi = \pm 1$ displays the self or anti-self duality, $\theta$ plays a two-fold role as the coupling constant and the (inverse) mass parameter for the dynamical fields.  The field strength of the basic potentials are,
\begin{eqnarray}
\label{PB20} F_{p+1}(A_p)=\partial_{[\mu_{0}}  A_{{\mu_1}
\ldots {\mu_p}]}\; .
\end{eqnarray}
Here the potentials play an active role in the duality transformations.  This shall be in contrast with the matter current, to be considered latter on -- although coupled to the potentials they are passive fields (spectators) in the duality mapping.
The equations of motion of the basic potentials $A_p$ and $B_q$
are, respectively,
\begin{eqnarray}
\label{PB30}
A_p &=& (-1)^{p}\,{\cal P}(\epsilon\partial)\, \frac {\chi\,\theta}{2} (\epsilon \partial B)_p \nonumber\\
B_q &=&  (-1)^{q}\, \frac{\chi\,\theta} 2 (\epsilon     \partial A)_q
\end{eqnarray}
satisfying the transversality constraint
\begin{eqnarray}
\label{PB40}
\partial\cdot M &=& 0 \:\:\:\:\: ; \:\:\:\:\:M = \left\{A_p ; B_q \right\}
\end{eqnarray}
identically.  
The parity property of the curl operator defined as,
\be\label{35}
 \left\langle B_{q} \epsilon \partial A_{p}\right\rangle
= {\cal P}(\epsilon\partial) \left\langle A_{p} \epsilon \partial B_{q}\right\rangle
\ee
is give by
\be
{\cal P}(\epsilon\partial) = (-1)^{(p+1)(q+1)}
\ee
Our notation goes as follows: $(\epsilon \partial B)_p = \epsilon_{p 1 q} \partial B^q \equiv \epsilon_{\mu_1\ldots\mu_p\alpha\nu_1\nu_q}\partial^\alpha B^{\nu_1\ldots\nu_q}$ is a p-form made out of the q-form $B_q$ and a generalized curl operator.  The equations (\ref{PB30}) constitute a set of first-order
coupled equations that can be combined into a decoupled second-order, massive, wave equations as
\begin{eqnarray}
\label{PB70}
\left(\Box + \frac{4}{p!q!\,\theta^2}\right) M&=& 0
\end{eqnarray}
whose mass depends crucially on the value of the coupling constant and the rank the potentials.

Next, we discuss the self-duality inherent to the above theory.  To this end we define a new derivative
duality operation by means of a set of star-variables as

\begin{eqnarray}
\label{PB90}
\mbox{}^*A_p &\equiv& (-1)^{p}\,{\cal P}(\epsilon\partial)\, \frac {\theta}{2} \left(\epsilon \partial B\right)_p \nonumber\\
\mbox{}^*B_q &\equiv&  (-1)^{q} \,\frac{\theta} 2 \left(\epsilon
\partial A\right)_q
\end{eqnarray}
With this definition we obtain, for the double duality operation, the relations
\begin{eqnarray}
\label{PB110}
\mbox{}^*\left(\mbox{}^* M\right) =  M \:\:\:\:\: ; \:\:\:\:\: M
= \left\{A_p ; B_q \right\}
\end{eqnarray}
after use of the equations of motion (\ref{PB70}).  This is important because it validates the
notion of self (or anti-self) duality
\begin{eqnarray}
\label{PB130}
\mbox{}^* M = \chi\, M \:\:\:\:\: ; \:\:\:\:\: M = \left\{A_p ; B_q \right\}
\end{eqnarray}
as a solution for the field equations, very much like the three-dimensional SD
model. However, this conceptualization of duality operation and self-duality in diverse dimensions is new.

Before we start the iterative procedure for the transformation of the $SD_{B\wedge F}$
model into a topological $B_q\wedge F_{p+1}$ model let us digress on the consequences of the self-duality relation (\ref{PB130}).
Notice first that under the usual gauge transformations of the potentials
\begin{eqnarray}
\label{PB140}
A_p & \to & A_p + F_p (\Lambda_{p-1}) \nonumber\\
B_q & \to & B_q + F_q (\Lambda_{q-1})
\end{eqnarray}
the fields strengths $F(A)$ and $F(B)$ are left invariant. Therefore, although the basic potentials are gauge dependent their duals, defined in (\ref{PB90}), are not. This situation parallels the three-dimensional case involving the Chern-Simons term which is the origin for the presence of a hidden (gauge) symmetry in the SD model of \cite{TPvN} while it is explicit in the topologically massive model of \cite{DJT}. Here too the $SD_{B\wedge F}$ model hides the gauge symmetry (\ref{PB140}) that is explicit in the $TM_{B\wedge F}$ model.
It will be shown the existence of such an intimate connection between the $SD_{B\wedge F}$ with a gauge invariant version through a dual transformation.  In the next section we shall discuss a dynamical gauge embedding procedure that will clearly produce an equivalent gauge invariant model.

\section{The Gauge Invariant $B_q\wedge F_{p+1}$ Theory}

In previous works  we have used the dynamical gauge embedding formalism to study dual equivalence in (2+1) \cite{IW,AINRW} and (3+1) dimensions \cite{BF} in diverse situations with models involving the presence of the topological Chern-Simons term and the ${B_2\wedge F_2}$ term, respectively. In this
section we extend that technique to study duality symmetry among models in diverse dimensions involving the presence of an extended topological $B_q\wedge F_{p+1}$ term. The minimal coupling
as both $A_p$ and $B_q$ tensor will be considered as well.

Our basic goal is to transform the symmetry (\ref{PB140}) that is hidden in the
Lagrangian (\ref{PB10})
into a local gauge symmetry by lifting the global parameter $\Lambda$ into its
 local form, i.e.,
$\Lambda\to \Lambda(x^\mu)$.
The method works by looking for an (weakly) equivalent description of the original
theory which may be obtained by adding a function $f(K_p , M_{q})$ to the Lagrangian
(\ref{PB10}). Here $K_{p}$ and $M_{p}$ are the Euler tensors, defined by the variation
\begin{eqnarray}
\label{PB180}
\delta {\cal L}_{SD}^{(0)} &=& K_p \,\delta A^p + M_q \,\delta B^q
\end{eqnarray}
whose kernels give the equations of motion for the $A_p$ and $B_q$ fields, respectively.
The minimal requirement for $f(K_p , M_q)$ is that it must be chosen such
that it vanishes on the space of solutions of (\ref{PB10}), viz. $f(0,0)=0$,
 so that the effective Lagrangian ${\cal L}_{eff}$
\begin{equation}
\label{PB190}
{\cal L}_{SD}^{(0)} \to {\cal L}_{eff}= {\cal L}_{SD}^{(0)} + f(K_p , M_q)
\end{equation}
is dynamically equivalent to ${\cal L}_{SD}^{(0)}$. To find
 the specific form of this function that also induces
a gauge symmetry into ${\cal L}_{SD}^{(0)}$ we work iteratively.
To this end we compute the variations (\ref{PB180}) of ${\cal L}_{SD}^{(0)}$
to find the Euler
tensors as
\begin{eqnarray}
\label{PB200}
K_p  &=& (-1)^{p+1} A_p + {\cal P}(\epsilon\partial)\, \frac {\chi\, \theta} 2 (\epsilon \partial B)_p   \nonumber\\
M_q &=&  (-1)^{q+1} B_q + \frac {\chi\, \theta} 2  (\epsilon \partial A)_q
\end{eqnarray}
and define the first-iterated Lagrangian as,
\begin{equation}
\label{PB210}
{\cal L}_{SD}^{(1)} = {\cal L}_{SD}^{(0)} - a_p K^p\, - b_q M^q
\end{equation}
with the Euler tensors being imposed as constraints and the new fields,
$a_p$ and $b_q$, to be identified with ancillary gauge fields, acting as a Lagrange multipliers.

The transformation properties of the auxiliary fields $a_p$ and $b_q$
accompanying the basic field
transformations (\ref{PB140}) is chosen so as to cancel the variation of
${\cal L}_{SD}^{(0)}$, which gives
\begin{eqnarray}
\label{PB220}
\delta a_p &=& \delta A_p \nonumber\\
\delta b_q &=& \delta B_q
\end{eqnarray}
A simple algebra then shows
\begin{eqnarray}
\label{PB230}
\delta {\cal L}_{SD}^{(1)} &=&  - a_p \,\delta K^p - b_q \,\delta M^p\nonumber\\
&=& \delta \left( \frac 12 (-1)^p  \,a_p^2 + \frac 12  (-1)^q b_q^2\right)
\end{eqnarray}
where we have used (\ref{PB140}) and (\ref{PB220}).  Because of (\ref{PB230}), the
second iterated Lagrangian is unambiguously defined as
\begin{equation}
\label{PB240}
{\cal L}_{SD}^{(2)} = {\cal L}_{SD}^{(1)} + \frac 12  (-1)^{p+1}  \,a_p^2 + \frac 12 (-1)^{q+1}  b_q^2
\end{equation}
that is automatically gauge invariant under the combined local transformation of the original set of fields
($A_p$, $B_q$) and the auxiliary fields ($a_p$, $b_q$).

We have therefore succeed in transforming the global $SD_{B\wedge F}$ theory into a locally
invariant gauge theory.  We may now take advantage of the Gaussian
character displayed by the auxiliary field to rewrite (\ref{PB240})
as an effective action depending only on the original variables ($A_p$, $B_q$).
To this end we use (\ref{PB240}) to solve for the fields $a_p$ and $b_{q}$ (call the
solutions $\bar a_p$ and $\bar b_{q}$ collectively by $\bar h_{\{p,q\}}$), and replace it back into (\ref{PB240}) to find
\begin{eqnarray}
\label{PB100}
{\cal L}_{eff}&=&{\cal L}_{SD}^{(2)}\mid_{h_{\{p,q\}} = \bar h_{\{p,q\}}} \nonumber\\
&=& {\cal L}_{SD}^{(0)} + \frac {1} {2} (-1)^p K_p^2  +  \frac {1} {2} (-1)^q
M_q^2
\end{eqnarray}
from which we identify the function $f(K_p , M_q)$ in (\ref{PB190}).
This dynamically modified action can be rewritten to give the $TM_{B\wedge F}$ theory,
\begin{equation}
\label{BI40}
{\cal L}_{eff} = \left(-1\right)^{q} \frac 18 \frac{p!}{(q+1)!} F^2_{q+1}(B_q)
                 +\left(-1\right)^{p} \frac 18 \frac{q!}{(p+1)!} F^2_{p+1}(A_p)
                 - \frac {\chi}{2\theta}   B_{q} \epsilon \partial A_{p} \; ,
\end{equation}
after the scaling $\theta\, A_p \to A_p$ and $\theta\, B_q \to B_q $ is performed. Notice the inversion of the coupling constant $\theta\to 1/\theta$ resulting from the duality mapping.
It becomes clear from the above derivation that the difference between these
two models is given by a function of the Euler tensors of the $SD_{B\wedge F}$ model that
vanishes over its space of solutions.  This establishes the
dynamical equivalence between the $SD_{B\wedge F}$ and the $TM_{B\wedge F}$ theory.

Once the duality mapping between the free theories has been established one is ready to consider the requirements for the existence of duality when the coupling with external matter current is included.  

The interacting Lagrangian now takes the form
\begin{equation}
\label{401}
{\cal L}_{min}^{(0)} = {\cal L}_{SD}^{(0)}  + e A_p J^p +  g B_q G^q
\end{equation}
with $e$ and $g$ being the strengths of the coupling with $A_p$
 and $B_{q}$, respectively.
The effective, gauge invariant action is obtained directly from
(\ref{PB100}) just operating the replacement
\begin{eqnarray}
\label{501}
K_p \to K_p^C &=&  K_p + e J_p \nonumber\\
M_q \to M_q^C &=& M_q + g  G_q
\end{eqnarray}
to produce
\begin{eqnarray}
\label{1000}
{\cal L}_{eff} &=& {\cal L}_{min}^{(0)}
 + \frac 1{2}(-1)^p (K_p^C)^2 + \frac 1{2}(-1)^q (M_q^C)^2
\end{eqnarray}
which, after some algebraic manipulation, gives
\begin{eqnarray}
\label{PB1102}
{\cal L}_{eff} &=& \left(-1\right)^{q} \frac 18 \frac{p!}{(q+1)!} F^2_{q+1}(B_q)
                 +\left(-1\right)^{p} \frac 18 \frac{q!}{(p+1)!} F^2_{p+1}(A_p)
                 - \frac {\chi}{2\theta}   B_{q} \epsilon \partial A_{p} \;  \nonumber\\
&&+   \frac {e^2}{2} (-1)^p\, J_p^2 + \frac{g^2}{2} (-1)^q  \, G_q^2
+ e   B_q \mbox{}^* J^q + g  A_p \mbox{}^* G^p
\end{eqnarray}
where  the dual currents are defined as
\begin{eqnarray}
\mbox{}^* J_q &=& (-1)^{p}\, \frac{\chi \theta}{2} \epsilon \partial J_p  \nonumber\\
\mbox{}^* G_p &=& (-1)^{q} \, {\cal P}(\epsilon \partial)\, \frac{\chi \theta}{2} \epsilon \partial G_q \, .
\end{eqnarray}
{}From this result it becomes clear the full action of the dual mapping over the active and passive fields involved in the transformation. Notice the exchange of the minimal coupling adopted in the SD sector into a non minimal, magnetic like interaction in the TM sector, including a swapping between the fields and currents and the presence of the current-current interaction for the matter sector.  This is needed to maintain the dynamics of the matter field (which here acts as an spectator field) unmodified \cite{GMdS}. To actually check for this fact depends crucially on the statistical character of the matter model adopted and will not be dealt with here.

Before concluding, let us digress on the connection of the present duality with the work of Quevedo and Trugenberger \cite{QT} on the condensation of electric and magnetic p-branes and the duality between Higgs and Julia-Toulouse mechanisms \cite{QT2}.
The investigation of mass generation for compact anti-symmetric tensors of arbitrary ranks, coupled to magnetic and electric topological defects, due to some condensation mechanism has been tackled in \cite{QT,QT2}.  An interesting duality between the Higgs and the Julia-Toulouse mechanisms for even dimensional spacetime was established where the Higgs phase is viewed as a coherent plasma of charged objects while the confinament phase is understood as a coherent plasma of monopoles.  When the dimension of the topological defects coincide the two mechanism are dual to each other.
In \cite{QT} and \cite{QT2} compact anti-symmetric field theories p-branes appear as topological defects of the original theory. While electric
(p-1)-branes coupled minimally with the original p-forms, the magnetic (d-p)-branes can be viewed as closed singularities (Dirac strings).  The opposite picture is valid for the $(d-p-1)$-form dual to the original tensor. The effective, low-energy field theory, is then valid outside these singularities. Topological defects condensation leads to drastic modifications of the infrared behavior of the original theory \cite{poly}. There is a new phase with a continous distribution of topological defects described by a low energy effective action - the condensation of topological defects gives rise to new low-energy modes representing the long-wavelength fluctuations about the homogeneous condensate. Quevedo and Trugenberger have shown that, in the presence of a magnetic defect described by a Dirac string (let us represent it by $\psi^{(0)}_{p}$), a massless abelian $(p-1)$-form $\phi_{p-1}^{(0)}$ interpolates into a massive $(p)$-form $\psi^{(m)}_{p}$ in the condensed phase of the magnetic defect. In this process, coined by them as Julia-Toulouse mechanism, the degrees of freedom of the abelian $(p-1)$-form are incorporated by the magnetic condensate to acquire a mass proportional to the density of the condensate,
\be
\phi_{p-1}^{(0)} \to  \psi_{p}^{(m)}=
\phi_{p-1}^{(0)} \oplus \psi_{p}^{(0)} \ee
This is quite distinct from the Higgs mechanism where the original $U(1)$ massless tensor $\phi^{(0)}_p$ acquires the degrees of freedom of the Higgs condensate, say $\Sigma_{p-1}^{(0)}$ to become massive,
\be
\phi^{(0)}_p  \to  \phi^{(m)}_p
= \phi^{(0)}_p \oplus \Sigma_{p-1}^{(0)} \ee
When the topological defects have the same dimensionality, Higgs and Julia-Toulouse phases are described by tensors of the same rank in this way establishing a duality between these two mechanisms \cite{QT}.

The result of the duality displayed in (\ref{BI40}) may be summarized by the following scheme,
\be
A^{(0)}_p  \to  A^{(m)}_p
 = A^{(0)}_p \oplus B_{q}^{(0)} \ee
if the ranks $p$ and $q$ of the massless fields $A_p^{(0)}$ and $B_q^{(0)}$ satisfy a massive duality condition: $p+q=d$.
We are now in position to compare the field contents of the present analysis with the mass generation coming from the Higgs and the Julia-Toulouse mechanism.  By inspection, we see,
\begin{itemize}
\item {Higgs/Soldering}
\be
\Sigma_{p-1}^{(0)} = \mbox{}^*\left(B_q\right)
\ee
\item{Julia-Toulouse/Soldering}
\be
\phi_{p-1}^{(0)} = \mbox{}^*\left(B_q\right)
\ee
\end{itemize}
where $*$ here is the massless duality operation, characterized by 
\be
\alpha_p = \mbox{}^*\beta_q
\ee
if $p+q+1=D-1=d$.  Therefore, in order to identify the fields we need the condition, $p-1 = q = d-p$ or, equivalently, $2p = d+1 = D$, that is the Quevedo-Trugenberger condition for the Higgs/Julia-Toulouse duality, to hold.  The field $B_q$ therefore interpolates between the original abelian form in the Julia-Toulouse condensation to the Higgs condensate in the Higgs mechanism.

\section{Conclusions}

In this work we studied dual equivalence of topological models, namely, between the $B_q\wedge F_{p+1}$ self-dual ($SD_{B\wedge F}$) and the $B_q\wedge F_{p+1}$ topologically massive ($TM_{B\wedge F}$) models, in diverse dimensions, using an iterative procedure of gauge embedding that produces the dual mapping. We defined a new derivative type of duality mapping, very much like the one adopted in the three-dimensional case and proved the self and antiself-duality property of the $SD_{B\wedge F}$ model, according to the relative sign of the topological term.
Working out the free case firstly, where the $A_p$ and $B_q$ fields participate actively in the dual transformation we observed, as expected, the traditional inversion in the coupling constant. The coupling to external matter current, whose fields act as spectators in the dual transformation, brought into the scene some new features. We mention the appearance of a Thirring like self-interaction term in the dualized theory, that had already been observed in the (2+1) case, as well as the shift from minimal to non minimal coupling. However, in this case we observed a swapping of the couplings from a tensor to another. This is a new result due to the presence of tensors of distinct ranks participating actively in the dual transformation.  The presence of these terms are demanded to maintain the equivalent dynamics in the matter sector in either representations of the duality. 

\vspace{.5cm}

\noindent ACKNOWLEDGMENTS: This work is partially supported by CNPq, CAPES,
FAPERJ and FUJB, Brazilian Research Agencies.  RM thanks the Physics
Department of UFRJ for the kind hospitality during the course of this
investigation.

\end{document}